\documentclass[a4paper,11pt]{article}

\usepackage{jinstpub} 

\title{\boldmath Test of streaming and triggered readout schemes for the TPEX Lead Tungstate Calorimeter}

\author[a,1,2]{I.~Fri\v{s}\v{c}i\'{c} \note{Corresponding author.}\note{Present address: Department of Physics, PMF, University of Zagreb, Croatia}}
\author[b]{E.~Cline}
\author[b,c]{J.C.~Bernauer}
\author[a]{D.K.~Hasell}
\author[a]{R.~Johnston}
\author[d]{I.~Lavrukhin}
\author[a]{S.~Lee}
\author[a]{P.~Moran}
\author[e]{U.~Schneekloth}

\affiliation[a]{Laboratory for Nuclear Science, Massachusetts Institute of Technology, Cambridge, MA, 02139, USA}
\affiliation[b]{Center for Frontiers in Nuclear Science, Department of Physics and Astronomy, Stony Brook University, Stony Brook, NY, 11794, USA}
\affiliation[c]{RIKEN BNL Research Center,  Brookhaven National Laboratory, Upton, NY, 11973, USA}
\affiliation[d]{Randall Laboratory of Physics, University of Michigan, Ann Arbor, MI 48109, USA}
\affiliation[e]{Deutsches Elektronen-Synchrotron DESY, Notkestr. 85, 22607 Hamburg, Germany}

\emailAdd{ifriscic@phy.hr,friscic@mit.edu}

\abstract{Tests of a prototype lead tungstate calorimeter were made over two weeks at the end of September, 2019, at the DESY II Test Beam Facility in Hamburg, Germany. The purpose of these tests was to gain experience with the construction, operation, and performance of a simple lead tungstate calorimeter, and also to compare a traditional triggered readout scheme with a streaming readout approach.  These tests are important for the proposed Two-Photon Exchange experiment at DESY \cite{TPEXproposal} and for work towards a future electromagnetic calorimeter that could be used in an Electron-Ion Collider detector. Details on 
the comparison of streaming and triggered readout schemes are presented here.}


\arxivnumber{2112.01606} 

\begin{document}
\maketitle
\flushbottom

\section{Introduction}
\label{sec:intro}

A prototype lead tungstate calorimeter was assembled at MIT and
shipped to the DESY laboratory in Hamburg, Germany where numerous tests were performed using the DESY II Test Beam Facility in TB24/1 area \cite{Diener:2018qap}.  These tests are part of a larger R\&D program to study numerous aspects of such calorimeters: the assembly and operation, readout and data acquisition.  Lead tungstate 
calorimeters have been extensively studied by the CMS 
experiment~\cite{Zhu:1996tt, Zhu:2004dg} and subsequent applications by the Bonn and Mainz groups at CEBAF~\cite{Neyret:1999tr} and for PANDA~\cite{Albrecht:2015zma}.  Our aim with this R\&D program was to gain experience in preparation for a possible Two-Photon Exchange eXperiment (TPEX) and for future electromagnetic calorimeters under consideration as part of an Electron-Ion Collider (EIC)  detector~\cite{ABDULKHALEK2022122447,ECCEEMCAL}.
 
The aim of the TPEX proposal is to determine the contribution of "hard" two-photon exchange in lepton-proton elastic scattering \cite{TPEXproposal}. Two-photon exchange is the leading explanation for the observed discrepancy between proton form factor ratio ($\mu_p G_E^p/G_M^p$) measurements using Rosenbluth separation and polarization transfer techniques. An in-depth review can be found in \cite{AFANASEV2017245}. Previous measurements \cite{Bernauer:2018big, OLYMPUS:2020dgl} observed
a small ($\sim$1\%) "hard" two-photon exchange contribution but were limited to relatively low $Q^2$ ($<2.1$~(GeV/$c$)$^2$) where the proton form factor discrepancy is not clear.  TPEX would make measurements at $Q^2$ up to $4.6$~(GeV/$c$)$^2$ where the discrepancy is more obvious. 
Measurements at higher momentum transfers would also be possible with some detector improvements.  

Lead-tungstate crystals were choosen for the TPEX calorimeter because of their high energy resolution, approximately 2\%,  that will be necessary to distinguish elastic scattered leptons from those produced in pion production.  Lead tungstate crystals have a density of 8.3 g/cm$^{3}$, and radiation length of $X_{0}$ = 0.8904 cm.  The energy resolution and light yield improves when the crystals are cooled. 

The excellent energy resolution of lead-tungstate crystals is also the reason behind their selection as part of the back angle electromagnetic calorimeter (EMCal) for an EIC detector~\cite{ABDULKHALEK2022122447}.  Nearly all EIC
physics processes require detecting the scattered electron.  Measuring the energy of back-angle scattered electrons is crucial for determining the kinematics of inclusive and semi-inclusive deep inelastic scattering.  It is likely 
that lead tungstate crystals will be used in the small angle region of the EMCal where the electron energies are small but the rates are high.

In addition to studying the performance of the lead tungstate
calorimeter two readout schemes were used in parallel for data 
acquisition. These were: a traditional triggered readout scheme and a streaming readout scheme. The term “streaming readout” refers to readout without a hardware-based trigger signal, where all detector information is recorded and the selection of events is performed at the software level, either in real-time or offline during data analysis. The aim for this was to directly compare 
the results from the two schemes to identify any issues that may need to be addressed in the future.  Streaming readout is not required for the TPEX measurements as the time between bunches will be large ($\sim$80~ms) leaving sufficient time to readout the detectors using a simple trigger based on the beam bunch clock. However, at the EIC streaming readout is widely accepted as the preferred and necessary approach to readout the large number channels from various detectors and present the data in a coordinated fashion to the clusters of computers that will select valid events and perform event
reconstruction~\cite{ECCEDAQ}.

The experimental setup and test procedures are briefly described below. This is followed by a description of the analysis including a comparison of the results from the two readout schemes.  Finally some conclusions are drawn and plans for future tests are presented.


\section{Experimental Setup}
\label{sec:Exp_setup}

The calorimeter used in the initial test run consisted of nine $2 \times 2\times 20$ cm$^3$ lead tungstate crystals from a set of 27 lead tungstate crystals~\cite{Horn:2019beh,Asaturyan:2021ese} generously loaned to us by Tanja Horn from the Catholic University of America. The crystals were produced by the Shanghai Institute of Ceramics of the Chinese Academy of Sciences, having a light yield >16 photoelectrons/MeV, and a decay time of 20 ns for the fast component. The light produced in each crystal was detected using Hamamatsu R1166 PMTs attached to one end of the crystal using optical coupling compound. The rest of the crystal surface was wrapped with one layer of white Tyvek (0.4 mm thick) to aid internal reflection and then wrapped with one layer of opaque aluminum foil (0.09 mm thick) to prevent light leakage and cross-talk between crystals. Passive voltage dividers for the PMTs were built according to the specifications provided in Hamamatsu manual. The crystal-PMT assembly was placed inside an anodized aluminum housing in $3\times 3$ configuration. SHV and BNC feedthroughs for high voltage and signal were mounted on opposing sides of the box.  Two 1/4" flexible copper tubes were pressed into machined tracks around the central part of the box. The copper pipes were connected to a cooling water loop of the laboratory at around 17 $^{\circ}$C to roughly stabilize the temperature. But no temperature dependent energy resolution was studied in this test.

\begin{figure}[ht]
\begin{center}
\includegraphics[scale=0.29]{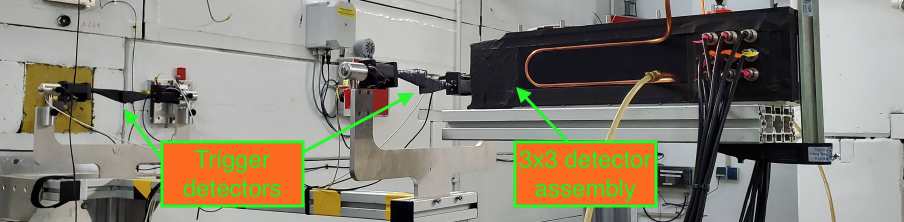}
\includegraphics[scale=0.115]{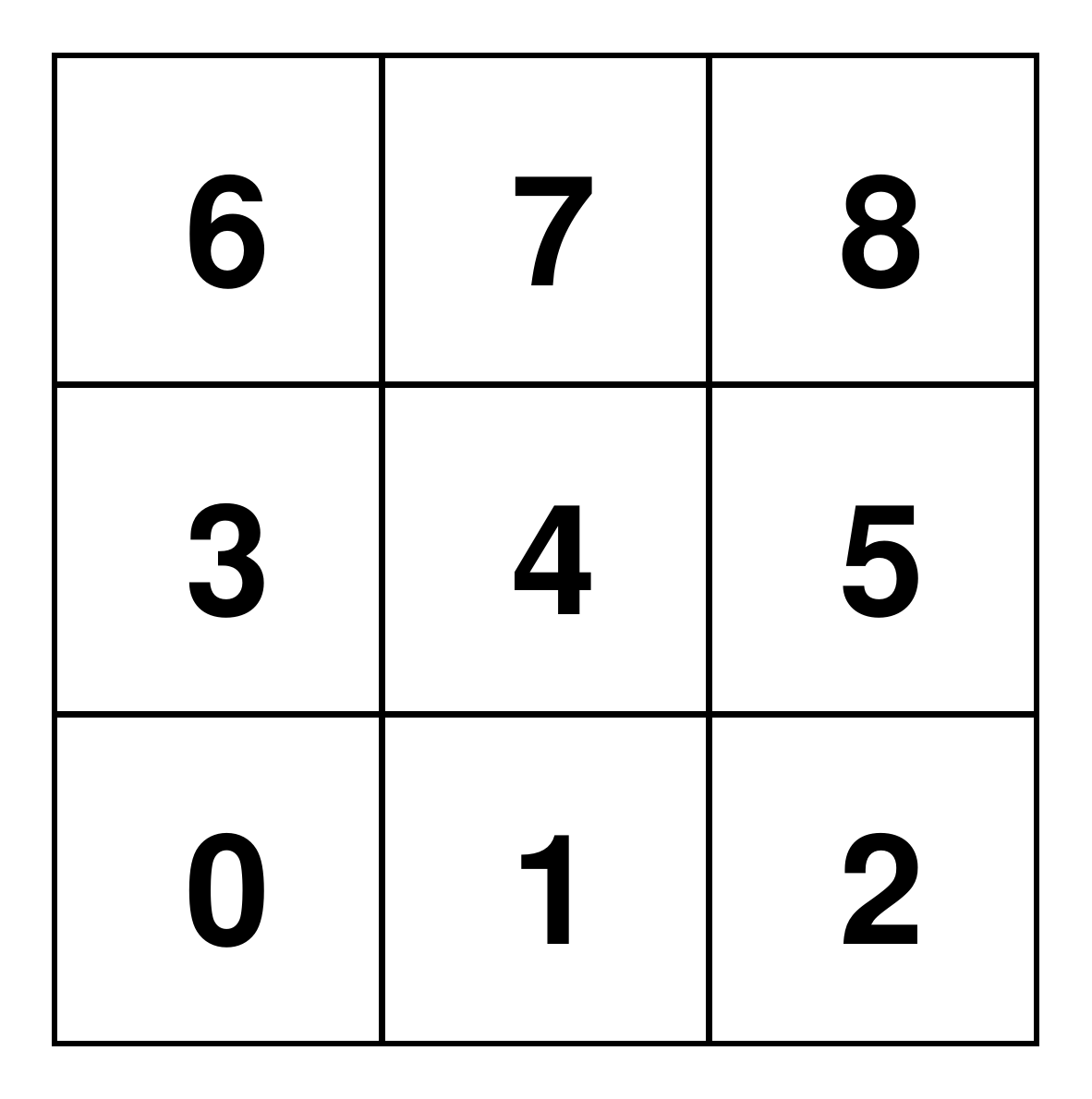}
\caption{\label{figIF1} (color online). Photo of 3x3 lead tungstate calorimeter prototype and trigger detectors used in the initial test run at DESY and the numbering scheme of the crystals.}
\end{center}
\end{figure}

The calorimeter assembly was mounted on a 2D translation table in front of four trigger detectors and the collimator opening for the electron beam, see Fig.~\ref{figIF1}.
Two overlapping scintillators were placed after the exit out of collimator and two scintillators with only overlapping edges were placed in front of calorimeter. The coincidence of all four trigger detectors generated the trigger signal. Vertical and horizontal laser beams could be used to indicate the beam position and determine translation table coordinates for the center of each crystal. When looking downstream into the front face of the calorimeter the crystals are numbered from 0 to 8 starting from the bottom left.

Power to the PMTs was provided by LeCroy 1461N modules housed in a LeCroy 1458 HV mainframe. Each LeCroy 1461N module has 12 independent SHV outputs providing up to -3 kV and 2.5 mA.

\begin{figure}[ht]
\begin{center}
\includegraphics[width=0.45\textwidth]{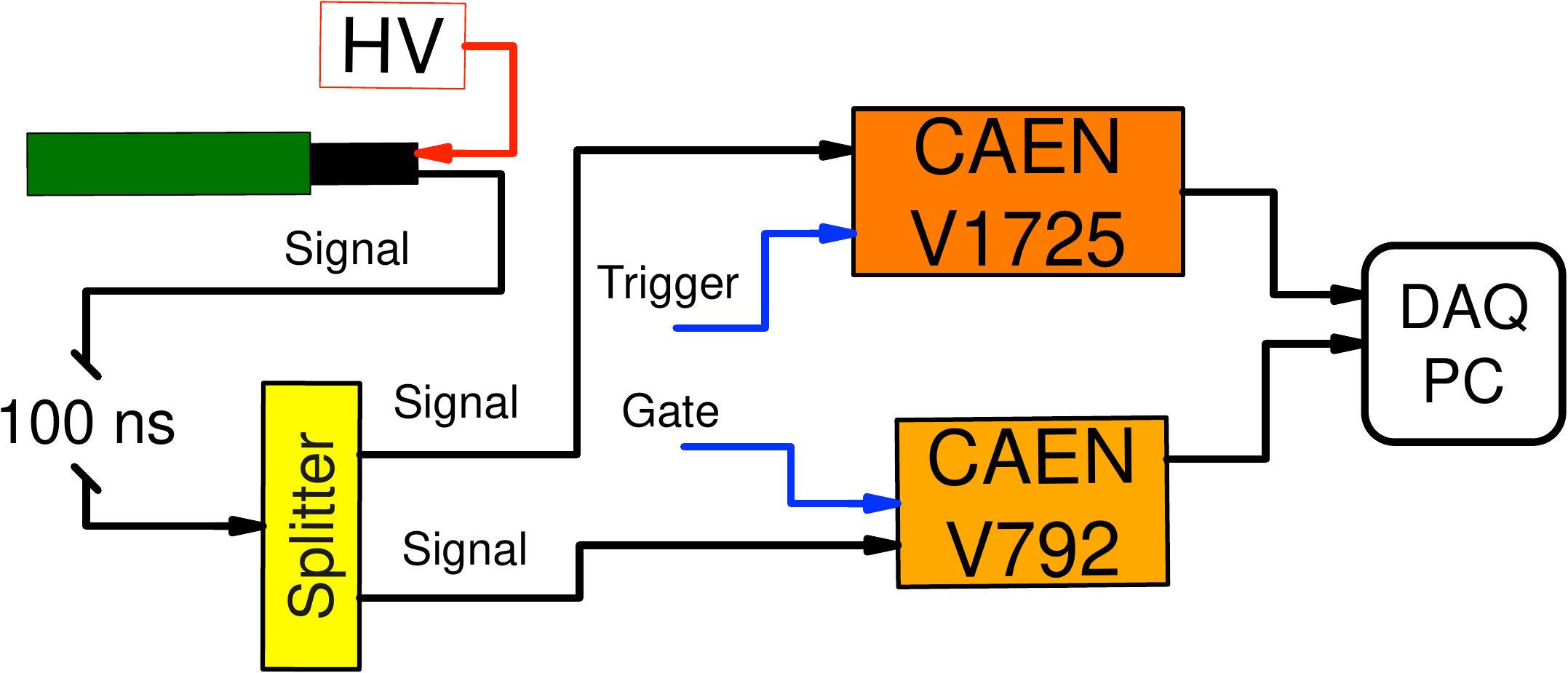}
\caption{(color online). The block diagram of the readout setup for a single calorimeter crystal. The signals from the trigger scintillators were sent to the coincidence module which generated a trigger signal. One output of this module was recorded by the CAEN V1725 digitizer and the other output was sent to the logic unit to generate a gate signal for the CAEN V792 QDC.
\label{fig:electronics_readout}}
\end{center}
\end{figure}

\begin{figure}[!ht]
\begin{center}
\includegraphics[width=0.45\textwidth]{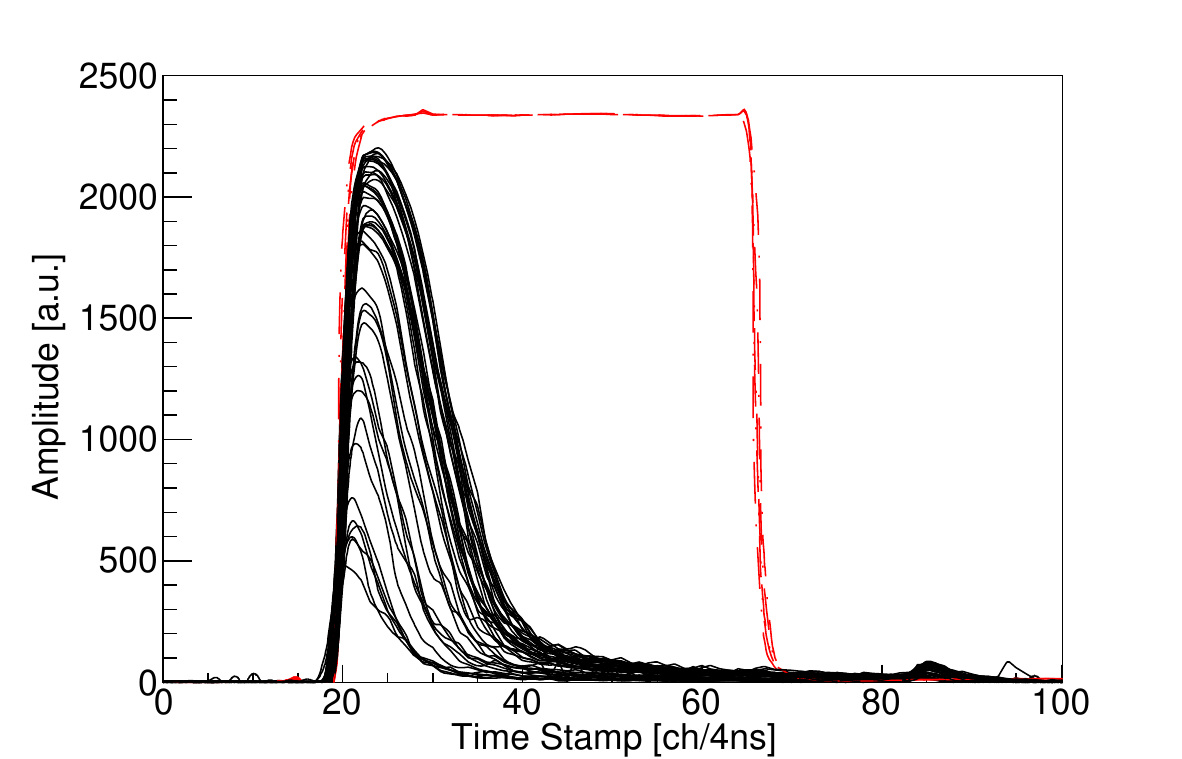}
\caption{(color online). A collection of waveforms of the signal from the central lead tungstate crystal (black, solid lines, showing 45 signals) and the trigger signal in channel 0 (red, dashed line, showing 5 signals, with the amplitude reduced by a factor of 3.0) as recorded by the CAEN V1725 digitizer. The waveforms shown here are baseline corrected and the sign was flipped to show signals with positive amplitudes. \label{fig:WaveForms}}
\end{center}
\end{figure}
A block diagram of the readout setup for a single calorimeter crystal is shown in Fig.~\ref{fig:electronics_readout}.
Signals from the PMTs were connected to a 50~ohm signal splitter. 
One splitter output was connected to a CAEN V792 channel in a classical triggered readout scheme and the other
was read out in streaming mode by a CAEN V1725 waveform digitizer.

The triggered readout signal was first delayed by 100 ns to allow it to fit into the window defined by the trigger logic and then read out using the CAEN V792 32 channel 12-bit QDC. For each trigger signal all channels were read out and the following information were saved to the PC: the channel number and the ADC converted value, and the PC time in seconds and in nanoseconds when the data for a particular trigger signal was recorded to the DAQ PC.

The CAEN V1725 has 14-bit resolution and a 250 MS/s sampling rate. Since this model of the digitizer has 8 channels, a decision was made to only read out crystals 1 to 7, and use channel 0 for the trigger signal. Baseline corrected waveforms recorded by the CAEN V1725 digitizer are shown in Fig.~\ref{fig:WaveForms}. For each detected signal, the following information was saved to disk: the waveform, the channel number, the online extracted energy from the full acquisition window width, the online extracted energy from the shorter acquisition window width, a time stamp (4 ns resolution), and the PC time in seconds and in nanoseconds when the data stream from the digitizer arrived at the DAQ PC. The energy extraction was performed in the signal shape analysis done by the CAEN DPP-PSD firmware in real-time  (Digital Pulse Processing for Charge Integration and Pulse Shape Discrimination) located in the FPGA of the V1725 board~\cite{Caen1}, by using the 424 ns (106 samples) long charge integration window.




\section{Analysis Methodology}
\label{sec:Tb_ana}

Before taking the test data each PMT's HV was adjusted to produce approximately the same amplitude of signal (gain matched) as measured in the CAEN V792. 
The beam position is static; therefore, the transition table was moved to place the center of each crystal into the beam
and data was collected with a 5.2~GeV electron beam. The high voltage for the selected crystal was adjusted to yield a peak in the QDC energy spectrum close to end of the QDC energy range with as little overflow as possible.

Data was then collected at four different beam energies: 2, 3, 4, and 5 GeV and with the translation table adjusted to center the beam on each crystal. 
This data was used to verify that the pedestal for each QDC channel does not change its position when changing the beam energy and in the analysis the pedestal was subtracted from the data. It also provided a measure of the linearity of the response versus energy.  

Horizontal and vertical scans ($\pm2$~cm) of the whole calorimeter were then made at the same four beam energies over the center crystal to study the response versus beam position. In this way the beam was centered in the middle of each crystal. Two different collimator size ($2\times2$~mm$^2$ and $8\times8$~mm$^2$) were used. The larger collimator was used for the rate comparison (see table \ref{Tab0}) and at the end to perform a fast stability check of the gain. All other data presented in this paper were taken by using the smaller collimator. In conjunction with the collimator, two lead absorbers, 1~cm and 2~cm thick, were placed in front of the calorimeter to perform energy loss studies. To complement this data, baseline data was taken without any lead absorbers.

\begin{figure}[!ht]
\begin{center}
\includegraphics[width=0.4\textwidth]{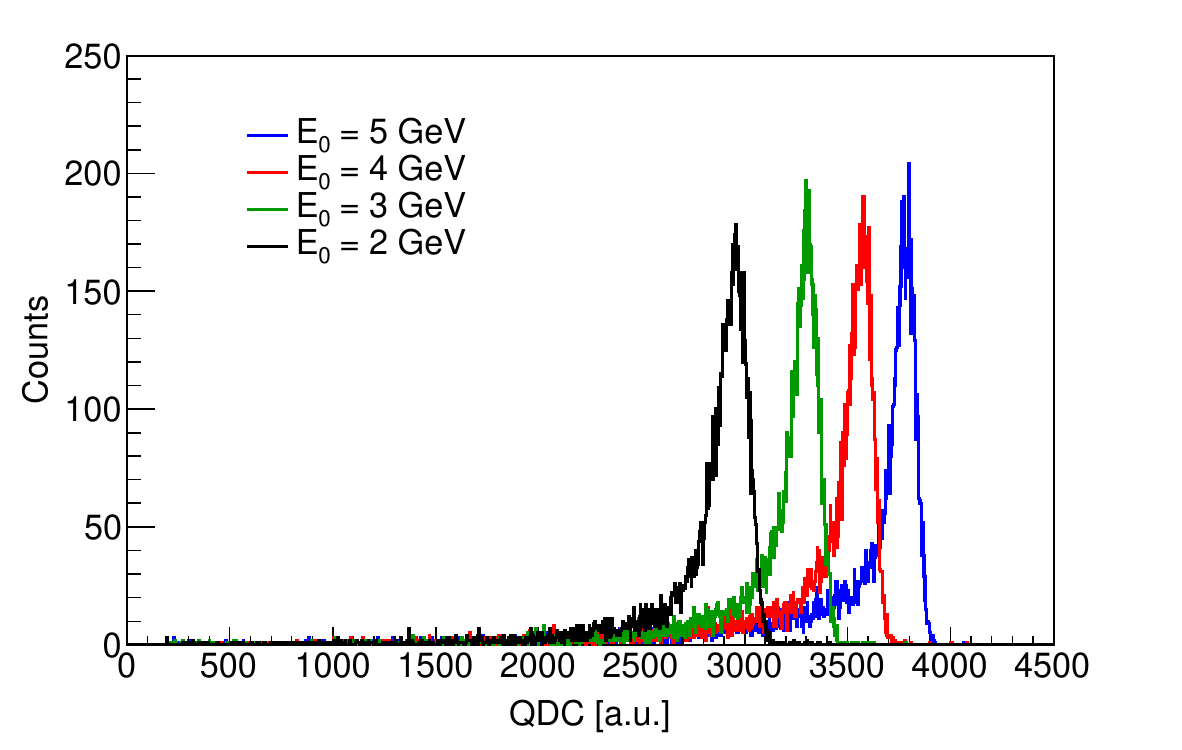}
\includegraphics[width=0.4\textwidth]{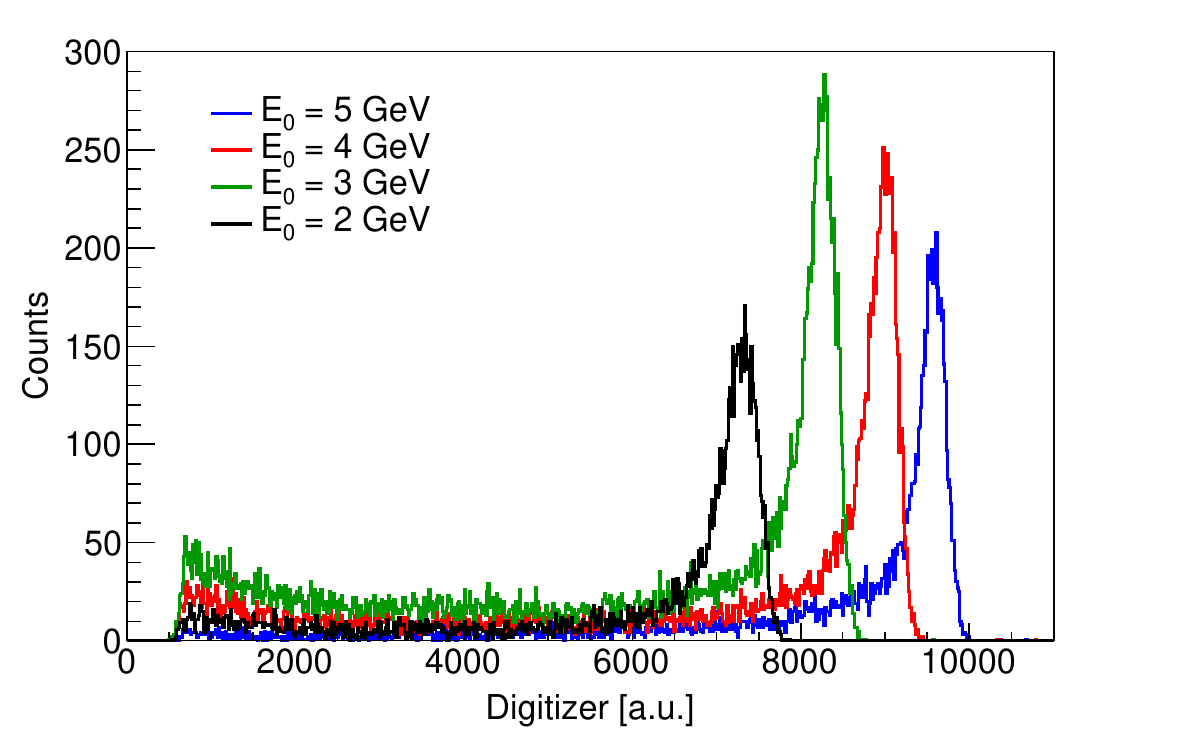}
\caption{(color online). Deposited energy in central crystal recorded by the QDC (left) and the digitizer (right) in central crystal. The events shown in the digitizer
spectra also required a coincidence with a trigger signal in channel 0 of the digitizer. \label{figIF2:spectra}}
\end{center}
\end{figure}

A plot of typical energy spectra for all four beam energies can be seen in Fig.~\ref{figIF2:spectra}. The QDC spectra (with pedestal subtraction) were normalized to 10k collected events at each energy. The digitizer spectra were normalized to 140k of total events recorded by the digitizer. Here, the peak sizes are affected by the beam current, which depends on beam energy~\cite{Diener:2018qap}. The beam current for 3 and 4 GeV is larger compared to 2 and 5 GeV, as can be seen in table \ref{Tab0}, which shows the detection rate of the central detector for two differenten collimator sizes when the beam was directed that detector.

\begin{table}[!ht]
\begin {center}
\caption{Detection rate when the beam is centered at the given detector.\label{Tab0}} 
\begin {tabular} {c|c|c|c|c|}
\cline{2-5}
                      & \multicolumn{2}{c|}{$2\times2$~mm$^2$ Collimator} & \multicolumn{2}{|c|}{$8\times8$~mm$^2$ Collimator} \\ \hline
\multicolumn{1}{|c|}{Beam energy (GeV)}  & QDC rate (Hz) & Digitizer rate (Hz) & QDC rate (Hz) & Digitizer rate (Hz)    \\ \hline
\multicolumn{1}{|c|}{2}   & 5.3  & 96.9  & 32.8  & 1293.2    \\ \hline
\multicolumn{1}{|c|}{3}   &  9.6  & 146.6  & 57.1 & 1936.8    \\ \hline
\multicolumn{1}{|c|}{4}  & 9.4  & 123.6  & 57.3  & 1428.1   \\ \hline
\multicolumn{1}{|c|}{5}   & 4.8  & 65.8  & 28.5  & 570.0    \\ \hline
\end {tabular} 
\end {center}
\end{table}


\begin{figure}[ht]
\begin{center}
\includegraphics[width=0.4\textwidth]{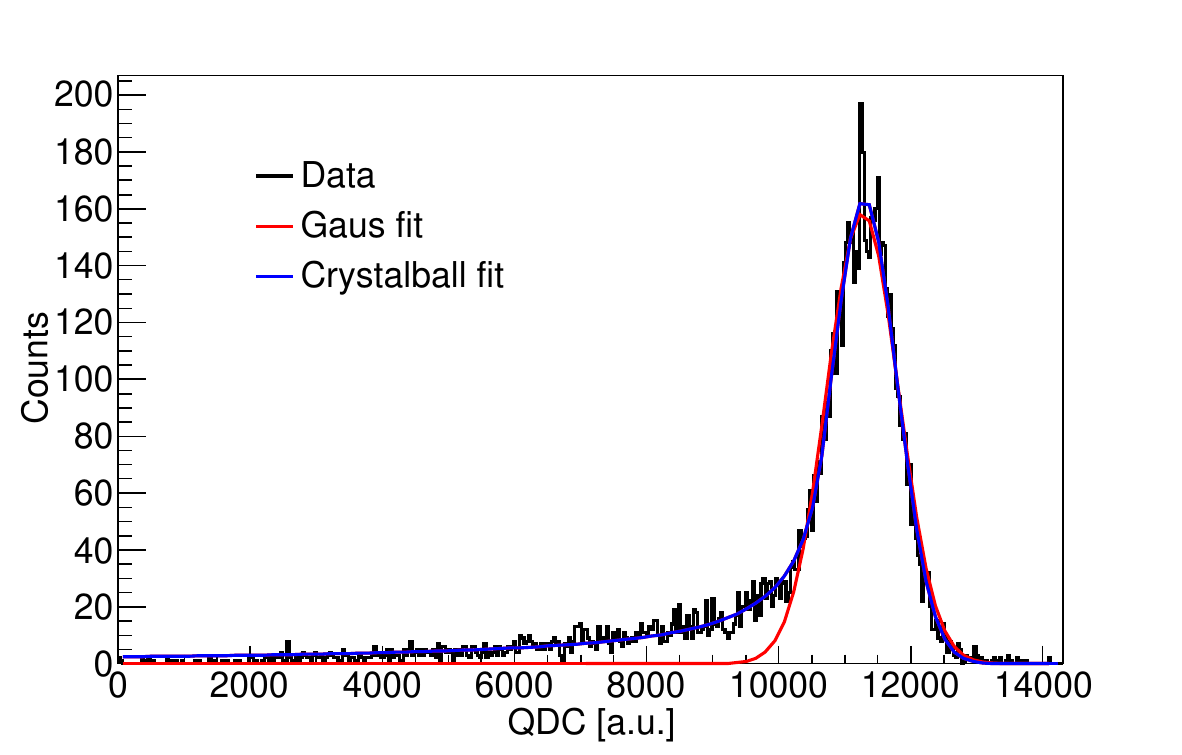}
\includegraphics[width=0.4\textwidth]{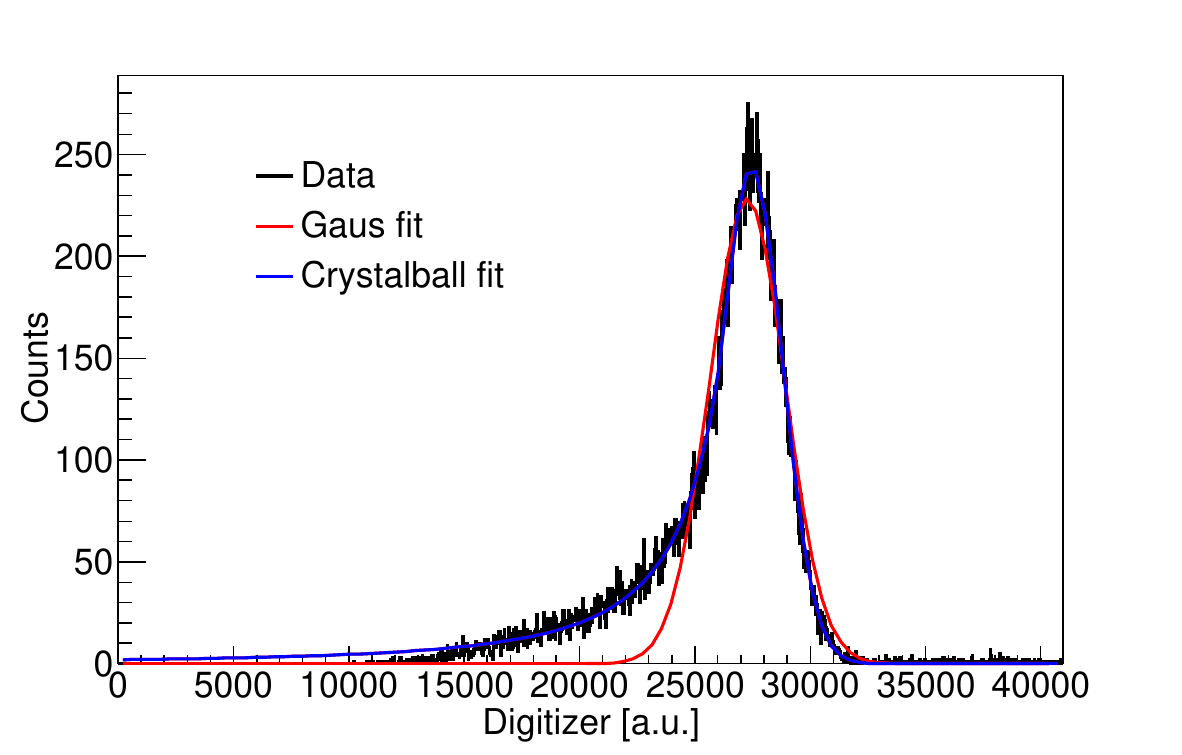}
\caption{(color online). Sum of energies deposited in all crystals (from 0 to 8) recorded by the QDC when 5 GeV electron beam was centered at crystal 4. The digitizer spectra is a sum of energy deposited in crystals from 1 to 7, which includes events without and also with
a coincidence with a trigger signal in channel 0 of the digitizer. In both cases, the pedestals were subtracted in all crystals. \label{figIF3:spectrasum}}
\end{center}
\end{figure}

When a high energy electron from the beam hits the crystal it will produce a particle shower which is not necessarily contained in a single crystal. In order to reconstruct the full energy of the particle one needs to sum the deposited
energy in all crystals for the given event. Figure \ref{figIF3:spectrasum} shows an example of this sum when the 5 GeV beam was directed at the 
central crystal. The ROOT~\cite{Bruna1997} functions ``gaus" and ``crystalball" \cite{OregliaPHD, GaiserPHD, SkwarnickiPHD} were used to extract the position of the peak. As can
be seen, the ``crystalball" function better fits the shape of the histogram than the ``gaus" function. 
 
\begin{figure}[!htb]
\begin{center}
\includegraphics[width=0.4\textwidth]{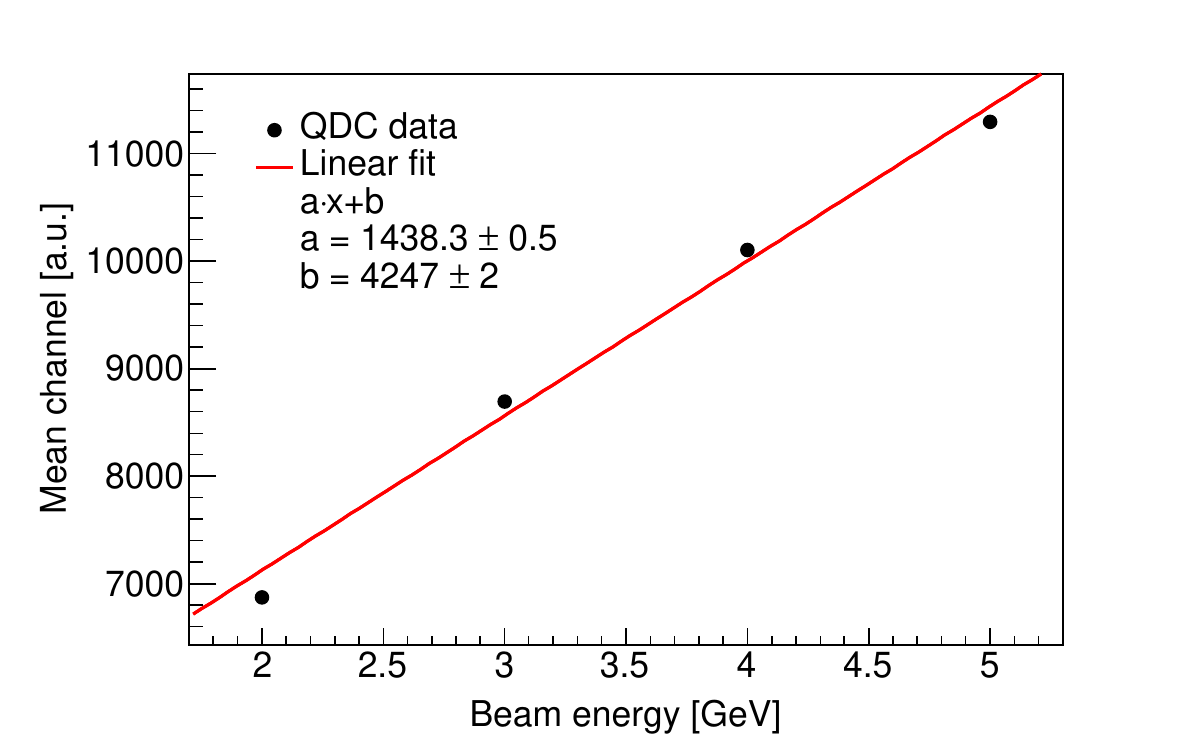}
\includegraphics[width=0.4\textwidth]{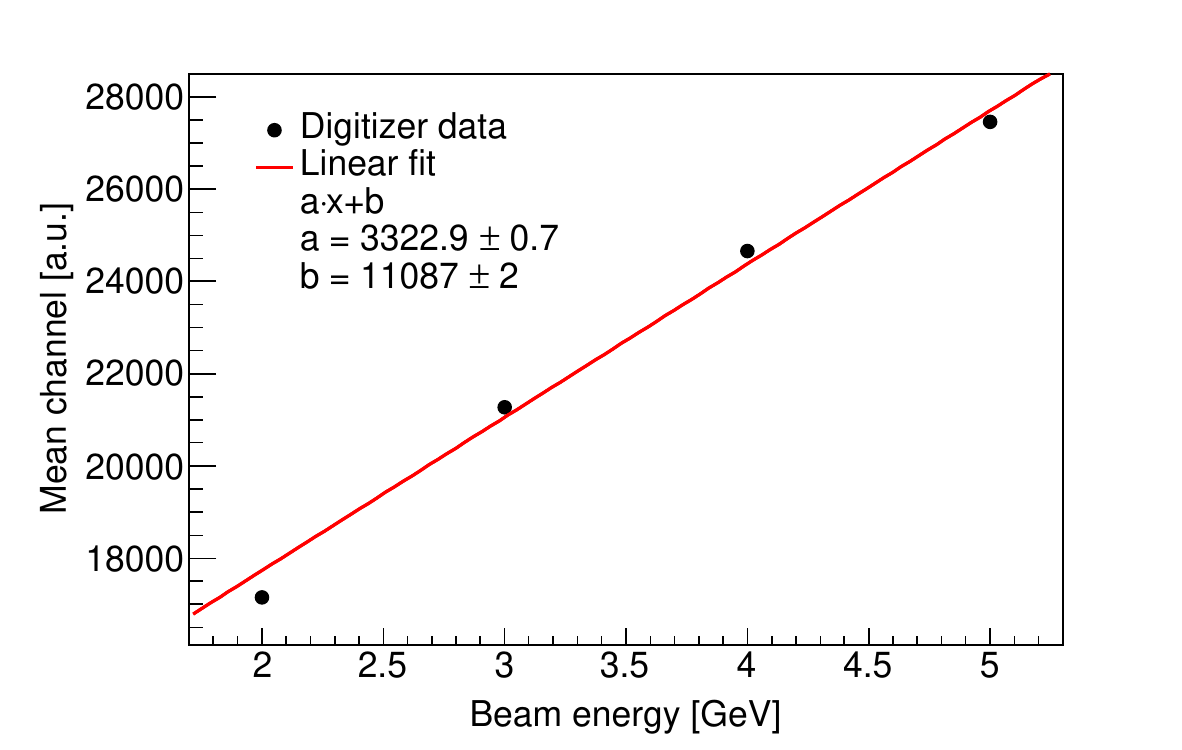}
\caption{(color online). Energy dependence of the peak position of the energy spectra sum in the QDC and the digitizer. The peak position is defined as the mean value extracted from the ``crystalball” fit. \label{figIF4:mean}}
\end{center}
\end{figure}

Positions of the sum peaks for all four energies were plotted against the beam energies in Fig.~\ref{figIF4:mean}, where one can observe a non-linearity which is increasing with the beam energy. 


\section{Streaming and Triggered Readout}
\label{sec:Tb_readout}

The DAQ system based on QDC electronics has a dead time of around 7.36 $\mu s$ and the one based on a digitizer is nearly dead time free, since the boards circular memory can store an event in a free buffer even if another buffer which stores an event is currently being read out \cite{CaenV1725}. The only situation when a dead time for the digitizer can occur is when all buffers are occupied and the board is in the ``full” condition and a new event is sent to the buffer~\cite{CaenV1725}, but this situation was avoided by reading out the memory when at least one event was recorded.

Due to QDC limitations, like dead time, requiring the trigger signal, etc., the QDC accepts a lower event rate than the digitizer. To account for this difference in recorded events, the data recorded by the digitizer without a corresponding trigger signal was removed from the analysis. Below we show our technique for removing these triggerless events for a 2 GeV run with the beam centered on the central crystal.

\begin{figure}[ht]
\begin{center}
\includegraphics[width=0.4\textwidth]{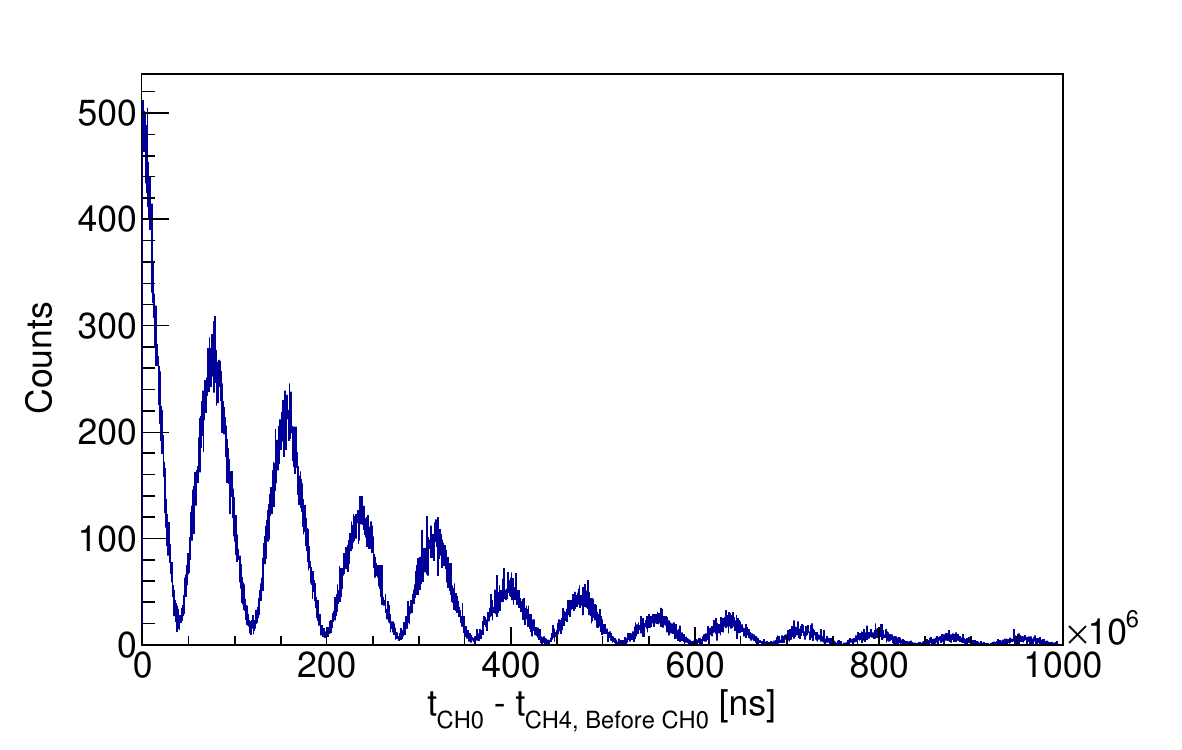}
\includegraphics[width=0.4\textwidth]{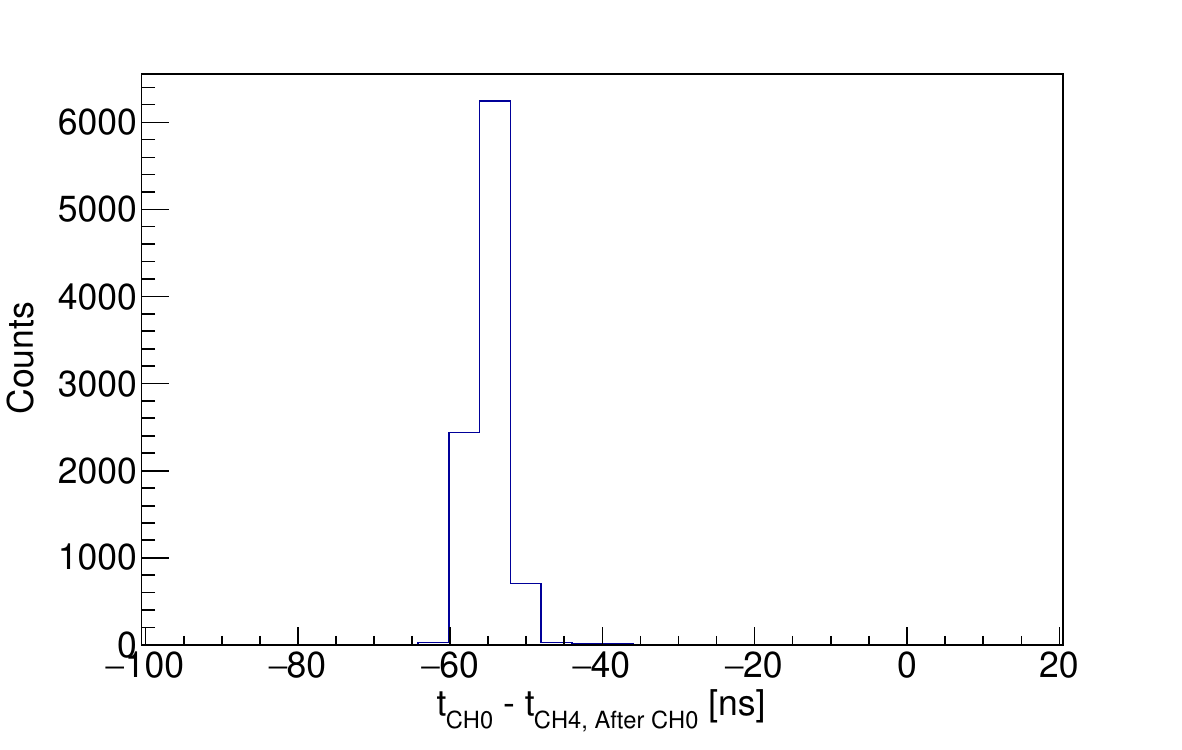}
\caption{(colour online). The time difference between channel 4 and the trigger signal in channel 0. The left panel shows the time difference for events in channel 4 that occurred before the trigger, and the right panel shows events that occured after the trigger signal. The regular structure that occurs in the left panel corresponds to the 12.5 Hz frequency when the beam enters the experimental hall. \label{figIF5:coin04}}
\end{center}
\end{figure}

The digitizer accepts all events above a threshold of 48 mV at the input dynamic range of 2 Vpp. These events must be correlated with the trigger signal. 
This was done by determining two time differences: for all events in a selected crystal between two subsequent events in the trigger (channel 0)
one time difference was formed with the first trigger signal and the other with the second trigger signal. In other words, we construct the time difference 
of events before the first trigger signal and the time difference of events after the trigger signal, as can be seen in Fig.~\ref{figIF5:coin04} for the central
crystal (channel 4). A similar timing offset was also found for other crystals. 

\begin{figure}[ht]
\begin{center}
\includegraphics[width=0.4\textwidth]{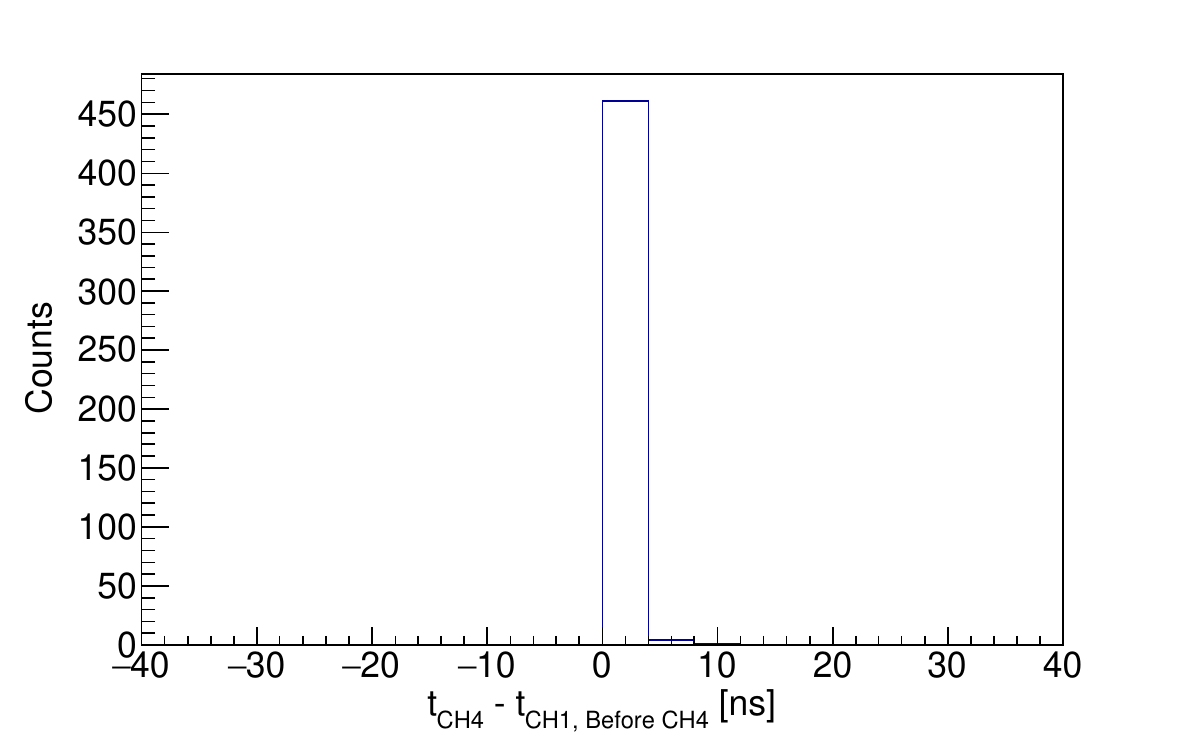}
\includegraphics[width=0.4\textwidth]{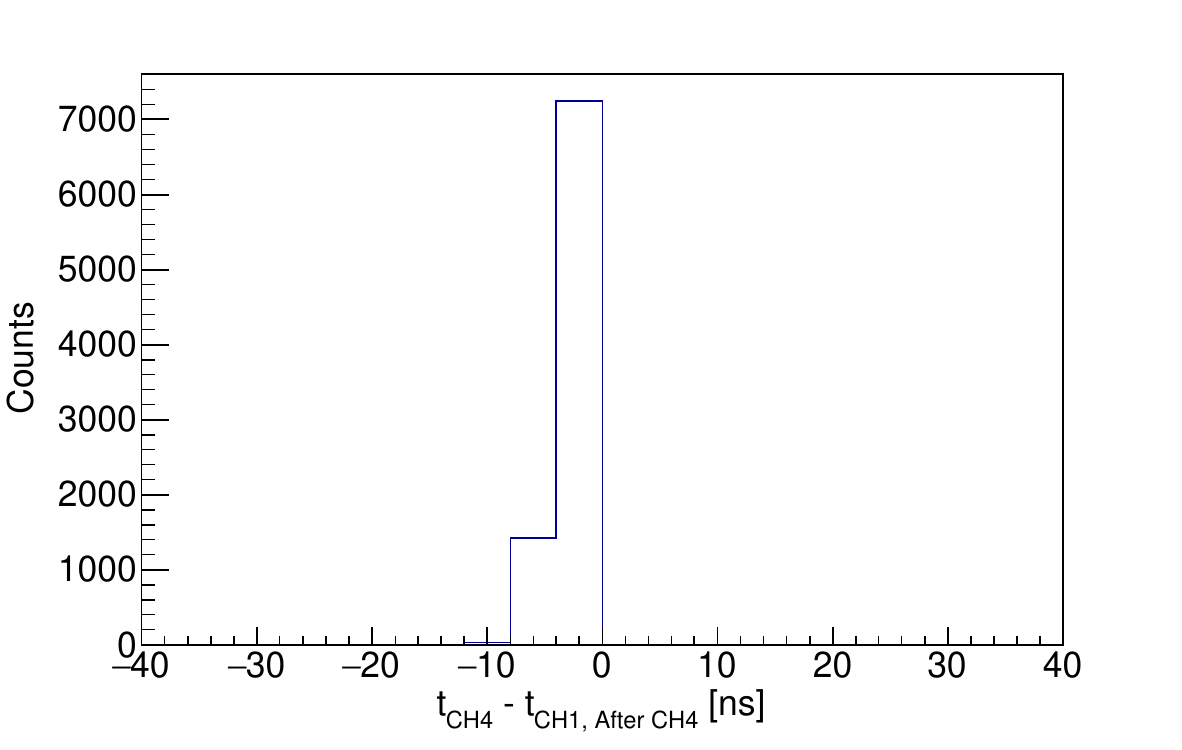}
\caption{Time difference between the channel 4 and channel 1 if the signal in channel 1 is recorded before (left) or after (right) the signal in channel 4. \label{figIF6:coin14}}
\end{center}
\end{figure}

The events forming a coincidence with the trigger were identified. The same procedure, as explained above, was used to determine
the coincidence between events from different crystals. Figure \ref{figIF6:coin14} clearly shows that these coincidences exist 
and the time offset between events in channel 4 and channel 1. Other channels had a similar offset from the trigger as channel 4.

\begin{figure}[ht]
\begin{center}
\includegraphics[width=0.4\textwidth]{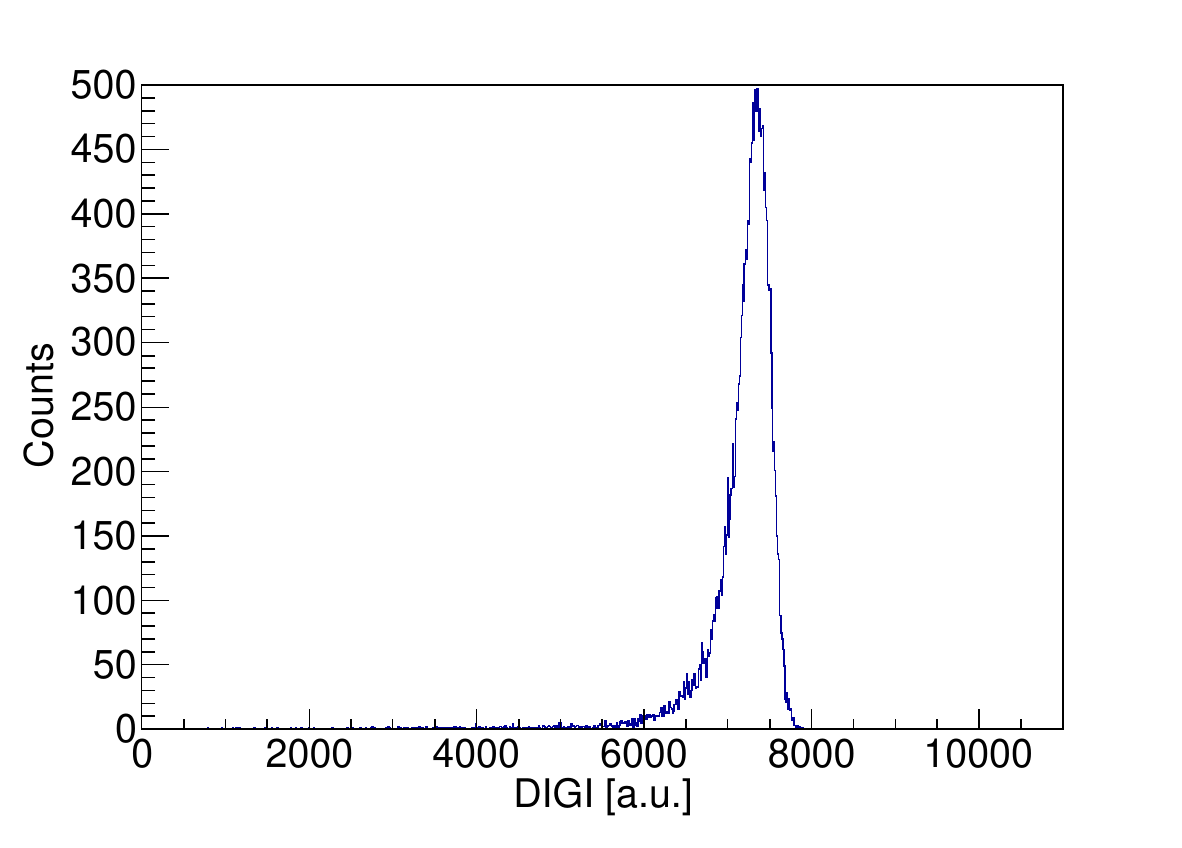}
\caption{Deposited energy in central detector (channel 4) for events which are in coincidence with at least 6 other events either in the digitizer channels or the trigger signal. 
\label{figIF7:digioff}}
\end{center}
\end{figure}

Using the determined time offsets for channel 4, the original data sample was searched for all events which are in agreement with
these offsets. Figure \ref{figIF7:digioff} shows all the selected events in the central crystal which are in coincidence
with at least 6 other events either in digitizer channels or the trigger signal. The number of such events (21452 events) is more than two times larger than the total number
of trigger events recorded by the QDC (9971 events) for this particular run. What demonstrates some limitations of the V792
performance at high rate.


In approximately 4\% of the events recorded by the digitizer, there was an error in the readout indicating that the digitizer conversion was not finished. This was a random occurrence during a run, so when selecting events in the digitizer that also had a trigger, we saw 4 \% fewer events than in the QDC.

Considering that the same data was measured with two different data acquisition schemes, the question arose if we can clearly identify events recorded in the QDC in the huge amount of data recorded by the digitizer (digitizer records $\sim$19 times more events in crystal targeted by the beam).


\begin{figure}[!hb]
\begin{center}
\includegraphics[width=0.4\textwidth]{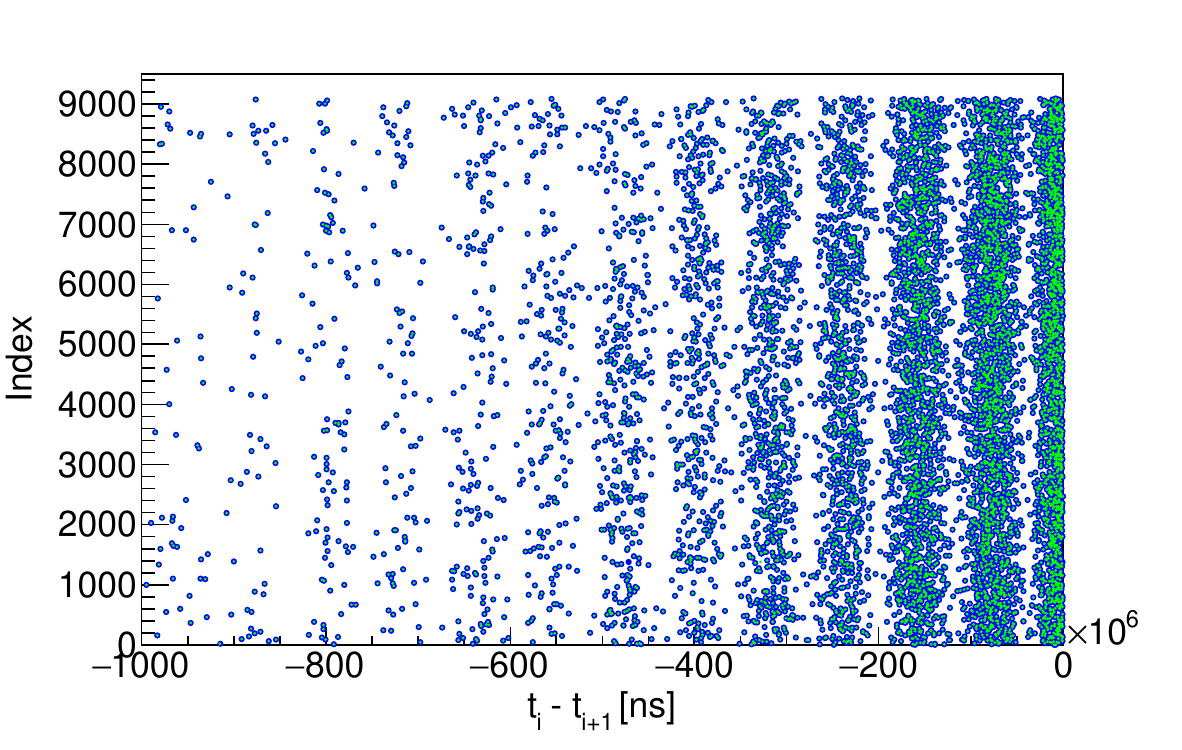}
\includegraphics[width=0.4\textwidth]{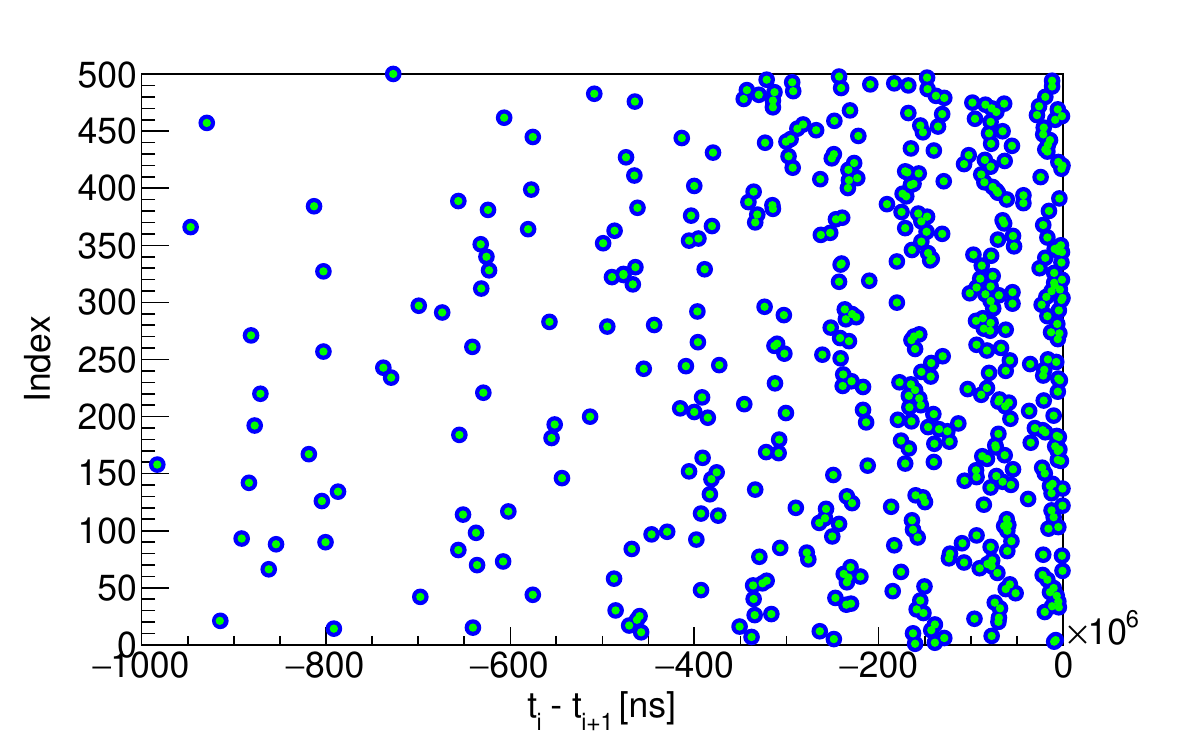}
\caption{(color online). Time interval between subsequent events in QDC (green points) and in digitizer (blue points). The left panel shows the full range in number of
time intervals. In order to show the good agreement between the QDC and the digitizer data, the size of the blue points was exaggerated and the right panel shows only the first 500 time intervals. \label{figIF8:timeDiff}}
\end{center}
\end{figure}

Both the QDC and the digitizer are read out by their own software which can be started independently from each other. The clock which defines the timing for the QDC data recording is totally asynchronous from the clock defining the 
timing for the digitizer data recording which is also asynchronous with the digitizer's own time stamps. 
This means that for each new run one would need to determine a new time offset between the clocks. In order to avoid this
complication, the solution is to calculate the time difference between the subsequent events in the QDC and the digitizer data. The
next step is to correct the 4\% fewer digitizer coincidence events compared to the QDC and one obtains data with 
excellent time alignment in both readout schemes as shown in Fig.~\ref{figIF8:timeDiff}. 

\begin{figure}[!h]
\begin{center}
\includegraphics[scale=0.4]{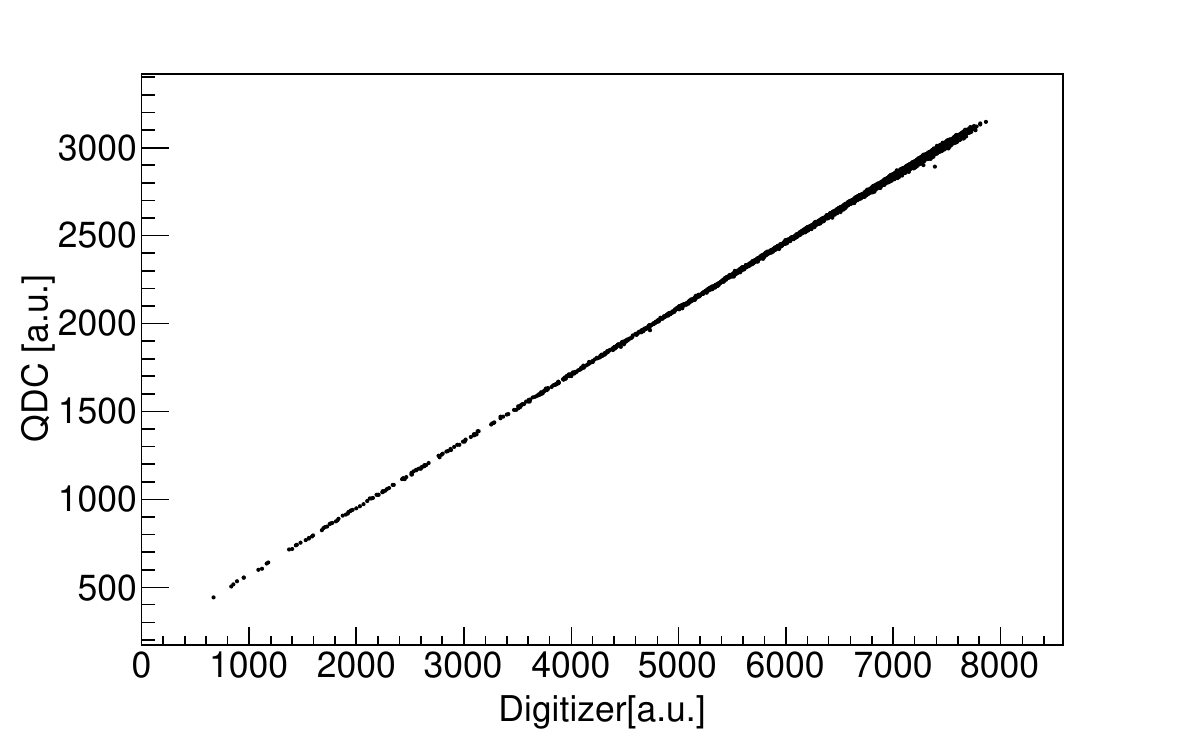}
\caption{The energy deposited in the digitizer versus QDC for events identified to belong to the  same signal recorded by the QDC and the digitizer for a single crystal. \label{figIF9:Energy}}
\end{center}
\end{figure}

An additional method of verification of the synchronization of the selected events, and to provide a cross-check of the readout schemes, is
to plot the energy recorded by the digitizer versus the energy recorded by the QDC. As can be seen in Fig.~\ref{figIF9:Energy} the energies recorded in two different readout schemes are correlated, meaning that we identified digitizer events which correspond to events recorded by the QDC on an event-by-event basis. 

\begin{figure}[!h]
\begin{center}
\includegraphics[scale=0.4]{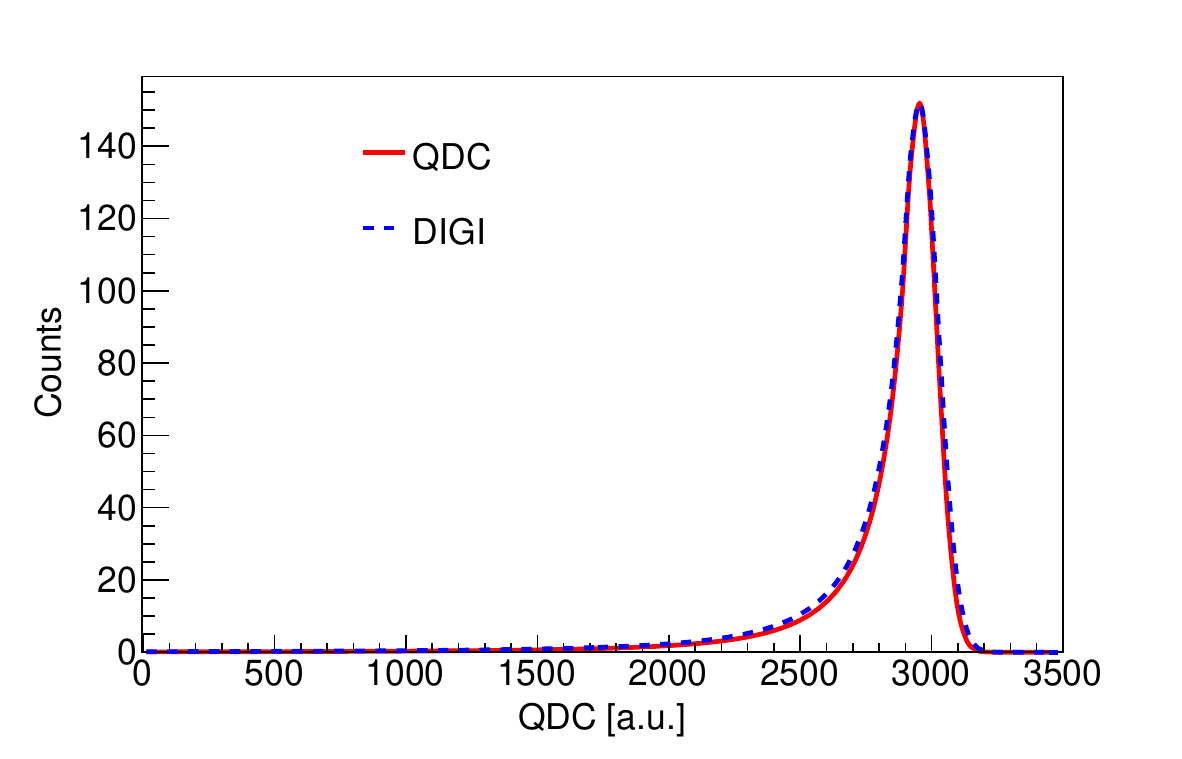}
\caption{(color online). Overlay of "crystalball" fits for the QDC and digitizer data shown in Fig. \ref{figIF9:Energy}. The original "crystalball" fit of the digitizer data was scaled down to the QDC data range. Therefore,  the amplitude $A$ and position $\bar{x}$ parameters were normalized with the QDC fit parameters, but the resolution $\sigma$ parameter was scaled down with the position normalization, other 2 parameters were not changed. \label{figIF9:crystallball}}
\end{center}
\end{figure} 

\begin{table}[!ht]
\caption{Parameters of the "crystalball" fit to the QDC and the digitizer data for the beam energy of 2 GeV. \label{Tab1}} 
\begin {center}
\begin {tabular} {l|c|c|c|c|c}
\hline
\multicolumn{6}{c}{"Crystalball" function fit parameters} \\ \hline
      & $A$    & $\bar{x}$  & $\sigma$    & $\alpha$   &    $n$ \\ \hline
QDC & 152.0$\pm$0.2 & 2952.7$\pm$0.2 & 65.3$\pm$0.2 & 0.681$\pm$0.007 & 2.87$\pm$ 0.07 \\ \hline 
DIGI & 154.5$\pm$0.2 & 7312.4$\pm$0.6 & 178.2$\pm$0.6 & 0.703$\pm$0.008 & 2.70$\pm$ 0.06 \\ \hline 
\end {tabular} 
\end {center}
\end{table}

Such selection of events is ideal to directly compare the two DAQ systems. For the CAEN V792 QDC the energy is given in terms of 12-bit ADC counts of charge integrated within the gate signal and which was later corrected for the pedestal. In the case of CAEN V1725 digitizer the signal shape is digitized by a 14-bit ADC and energy extraction in terms of counts is performed in real-time by the CAEN DPP-PSD firmware (Digital Pulse Processing for Charge Integration and Pulse Shape Discrimination) located in the FPGA of the V1725 board. The spectra recorded by the two DAQ systems were fitted with a ``crystalball" function, the parameters of the fits are given in table \ref{Tab1}. The relative resolution obtained is $\sigma/\bar{x} = 2.2\%$ for the QDC and $\sigma/\bar{x} = 2.4\%$ for the digitizer. In order to produce an overlay of these two fits shown in figure \ref{figIF9:crystallball}, the fit of the digitizer was rescaled to the QDC range. This was done by normalizing the amplitude $A$ and position $\bar{x}$ of the digitizer peak with the QDC data. The resolution $\sigma$ was normalized with the same factor as the position $\bar{x}$ and the other two parameters were unchanged. As shown in figure \ref{figIF9:crystallball}, both fits are very similar, they only differ in their widths, which is due to $\sim$10\% larger relative resolutions obtained from the digitizer data. This difference is rather small, and could later be improved in future firmware upgrades or by trying other signal shape analysis methods, but this is beyond the scope of this paper.

\begin{figure}[!htb] 
\begin{center}
\includegraphics[scale=0.6]{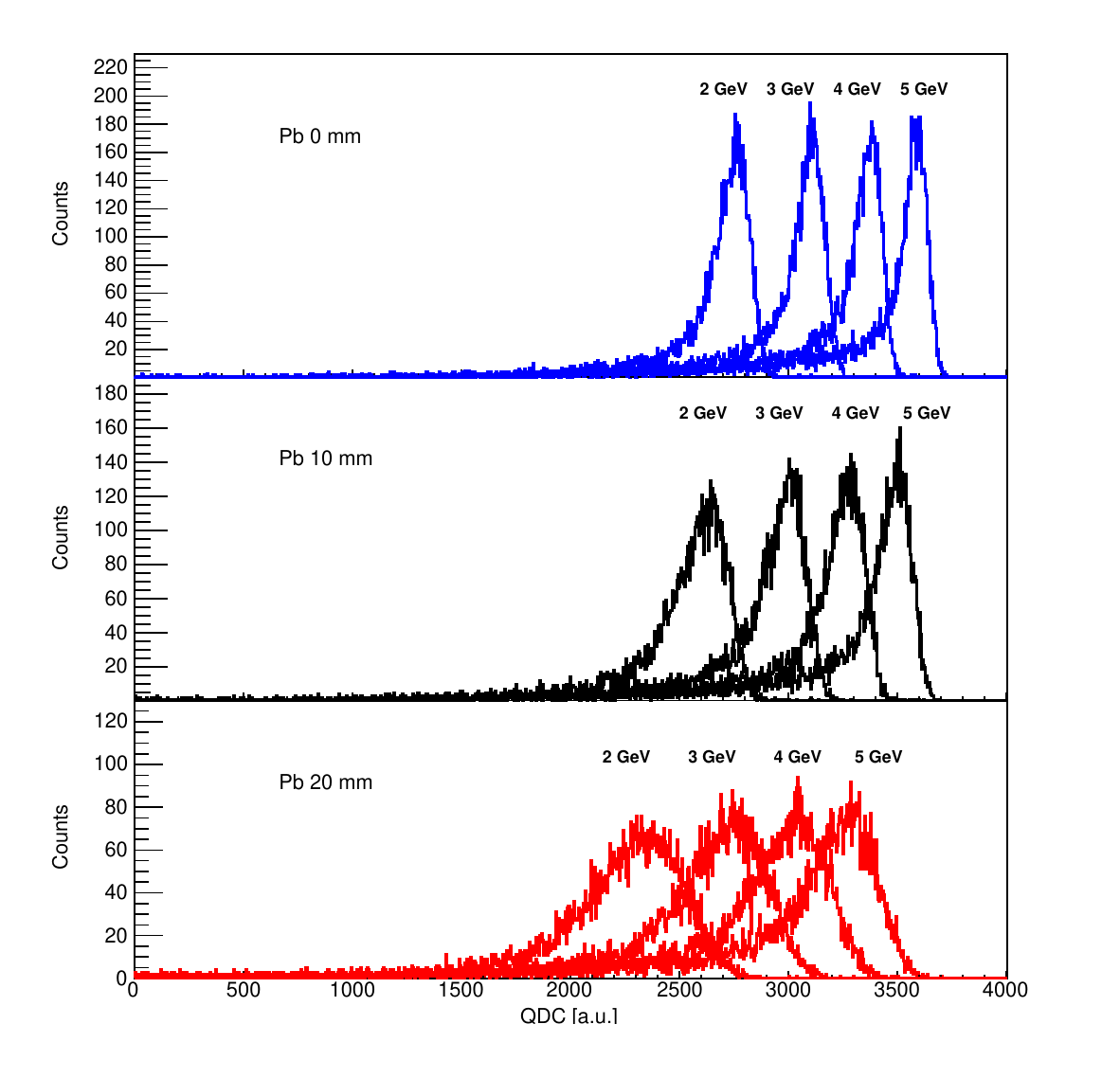}
\caption{(color online). Distribution of energy deposited in the central crystal for the case without a lead absorber (upper panel) at four beam energies and evolution of corresponding peaks for measurements with lead absorber in front of the calorimeter having thickness of 10 mm (middle panel) and 20 mm (lower panel).  \label{figIF10:PbAll}}
\end{center}
\end{figure} 
In order to further study the detector response two sets of data with four beam energies have been taken with a lead absorber being placed in front of the calorimeter. The lead could be used as a preshower material, and also as a way to reduce the background induced by M\o ller and Bhabha scattering in the TPEX experiment. One data set was taken with a lead absorber having thickness of 10 mm and the other data set with 20 mm. In both cases the electron beam was centered at the central crystal. Figure \ref{figIF10:PbAll} shows energy deposited in the central crystal for these two data sets compared with the measurement without the lead absorber.


\section{Conclusions}
\label{sec:Tb_conc}

A prototype, 3$\times$3, lead tungstate calorimeter was tested at the DESY II Test Beam Facility with an electron beam having energy 2, 3, 4 and 5 GeV. 
The signals from PMTs were split with one signal read out using the data acquisition system based on a QDC and a signal provided by trigger scintillators and the other signal read out by signal shape digitizers (based on a streaming readout scheme). 
Both readout schemes consistently showed slight non-linearity when comparing the beam energy dependence and the position of the total deposited energy peak in calorimeter.  This indicates that part of the energy from the primary electron beam is being lost between the individual crystals when it is being converted into the shower of secondary particles. The amount of energy loss tends to be larger for the higher beam energy.

We demonstrated that in the same time the digitizer collects more data compared to a readout based on a QDC. This was expected, since the QDC requires the trigger signal and also has a dead time and the readout based on the digitizer does not have such limitations. Furthermore, by using the time difference between subsequent events, we managed to identify events in a large sample collected by the digitizer which are synchronized with the events collected by the triggered QDC.

For all four beam energies, data was also taken with two lead absorbers being placed in front of the calorimeter. One absorber had a thickness of 10 mm and the other 20 mm.  When overlaying the data for all beam energies and absorber thicknesses one can clearly observe a broadening and a movement of the recorded peak towards lower QDC channels if one increases the absorber thickness. 


For the future, we plan to perform new tests at the DESY II Test Beam Facility with an improved version of the lead tungstate calorimeter. First, we will increase the number of crystals and build a 5$\times$5 lead tungstate calorimeter. Second, by using a thinner wrapping material the gap between individual crystals will be smaller, hence, undetected energy from the particle shower will be further reduced. The new test measurements will be performed in TB24 area, where we will have a detailed knowledge of the beam path from the beam extraction point to the calorimeter that will help us to also incorporate the Geant4 simulation in the analysis.  We also will compare the data acquisition with the FADC250 system from JLab.

Of particular interest for future tests is the possibility of obtaining some high-density, scintillating glass as an alternative to the lead tungstate crystals.  This work is being developed by Tanja Horn at Catholic University of America. The density is approximately 15\% less than lead tungstate and the resolution slightly poorer but it is not sensitive to temperature and can run at room temperature. A big advantage is that the cost is expected to be 5--10 times less.


\section{Acknowledgements}
\label{sec:Acknow}

The measurements leading to these results have been performed at the
Test Beam Facility at DESY Hamburg (Germany), a member of the 
Helmholtz Association (HGF). The setup and operation of these tests 
were extremely easy and enabled us to achieve everything we needed in 
a relatively short time thanks to the help of the test beam coordinators:
Ralf Diener, Norbert Meyners, and Marcel Stanitzki. We are very grateful to Tanja Horn of Catholic University of America for loaning us the lead tungstate crystals used in this test. Obviously the tests would not have been possible without her assistance. We also acknowledge the generous support from several funding agencies without which none of this would be possible. These include:
DOE Office of Science grant DE-FG02-94ER4081, PIER Hamburg-MIT/BOS Seed Project PHM-2019-04, National Science Foundation under Grant 2012114 and National Science Foundation under Grant 2110229.

\bibliographystyle{JHEP}
\bibliography{references.bib} 
\end{document}